\documentclass[journal,onecolumn,letterpaper,10pt]{IEEEtran}
\usepackage{color}
\usepackage[center, font={footnotesize}]{caption}
\usepackage{array}
\usepackage{cite}
\usepackage{multirow}
\usepackage{stmaryrd}
\usepackage{setspace}
\usepackage{amssymb}
\usepackage[cmex10]{amsmath}
\usepackage{amsthm}
\usepackage{graphicx}
\usepackage{epsfig}
\usepackage{verbatim}
\newsavebox{\subfigure}
\newsavebox{\theorembox}
\newsavebox{\lemmabox}
\newsavebox{\corollarybox}
\newsavebox{\propositionbox}
\newsavebox{\examplebox}
\newsavebox{\conjecturebox}
\newsavebox{\algbox}
\newsavebox{\qbox}
\newsavebox{\problembox}
\newsavebox{\definitionbox}
\newsavebox{\assumptionbox}
\newsavebox{\hypothesisbox}
\newsavebox{\factbox}
\newsavebox{\remarkbox}
\savebox{\theorembox}{\noindent\bf Theorem}
\savebox{\lemmabox}{\noindent\bf Lemma}
\savebox{\corollarybox}{\noindent\bf Corollary}
\savebox{\propositionbox}{\noindent\bf Proposition}
\savebox{\examplebox}{\noindent\bf Example}
\savebox{\conjecturebox}{\noindent\bf Conjecture}
\savebox{\algbox}{\noindent\bf Algorithm}
\savebox{\qbox}{\noindent\bf Question}
\savebox{\definitionbox}{\noindent\bf Definition}
\savebox{\problembox}{\noindent\bf Problem}
\savebox{\assumptionbox}{\noindent\bf Assumption}
\savebox{\hypothesisbox}{\noindent\bf Hypothesis}
\savebox{\factbox}{\noindent\bf Fact}
\savebox{\remarkbox}{\noindent\bf Remark}
\newtheorem{theorem}{\usebox{\theorembox}}

\newtheorem{corollary}{\usebox{\corollarybox}}

\newtheorem*{fact}{\usebox{\factbox}}

\begin{document}
\title{\huge An Efficient Approach Toward the Asymptotic Analysis of Node-Based Recovery Algorithms in Compressed Sensing}
\author{Yaser Eftekhari, Amir H. Banihashemi, Ioannis Lambadaris\\
			Carleton University, Department of Systems and Computer Engineering, Ottawa, ON, Canada}
\maketitle
\thispagestyle{empty}
%\newpage
%\linenumbers
%\linenumbersep 30pt\relax
%\linenumbersep 30pt
\begin{abstract}
In this paper, we propose a general framework for the asymptotic analysis of node-based verification-based algorithms. In our analysis we tend the signal length $\boldsymbol{n}$ to infinity. We also let the number of non-zero elements of the signal $\boldsymbol{k}$ scale linearly with $\boldsymbol{n}$. Using the proposed framework, we study the asymptotic behavior of the recovery algorithms over random sparse matrices (graphs) in the context of compressive sensing.
Our analysis shows that there exists a success threshold on the density ratio $\boldsymbol{k/n}$, before which the recovery algorithms are successful, and beyond which they fail. This threshold is a function of both the graph and the recovery algorithm.  
We also demonstrate that there is a good agreement between the asymptotic behavior of recovery algorithms and finite length simulations for moderately large values of $\boldsymbol{n}$.
\end{abstract}
%%%%%%%%%%%%%%%%%%%%%%%%%%%%%%%%%%%%%%%%%%%%%%%%%%%%%%%%%%%%%%%%%%%%%%%%%%%%%%%%%%%%%%%%%%%%%%%%%%%%%%%%%%%%%%%%%%%
%%%%%%%%%%%%%%%%%%%%%%%%%%%%%%%%%%%%%%%%%%%%%%%%%%%%%%%%%%%%%%%%%%%%%%%%%%%%%%%%%%%%%%%%%%%%%%%%%%%%%%%%%%%%%%%%%%%
\section{Introduction}
Compressive sensing was introduced with the idea to represent a signal $\underline{V}\in\mathbb{R}^n$ having $k$ non-zero elements with measurements $\underline{C}\in\mathbb{R}^m$, where $k<m\ll n$ and yet be able to recover the original signal $\underline{V}$ back \cite{D06,CRTFeb06}. In the measuring process, also referred to as \textit{encoding}, signal elements are mapped to measurements through a linear transformation represented by the matrix multiplication $\underline{C} = \underline{V}\textbf{G}$, where the matrix $\textbf{G}\in \mathbb{R}^{n\times m}$ is referred to as the \emph{sensing matrix}. This linear mapping can also be characterized by a bipartite graph \cite{XH07}, referred to as the \textit{sensing graph}.

In the recovery process, also referred to as \textit{decoding}, based on the knowledge of the measurements and sensing matrix, we try to estimate the original signal. For given $n,m,$ and $k$, a decoder is called \textit{successful} if it recovers the original signal thoroughly. Two performance measures namely, density ratio $\gamma \triangleq k/n$ and oversampling ratio $r_o \triangleq m/k$ are used in order to measure and compare the performance of the recovery algorithms in the context of compressive sensing\footnote{For successful decoding clearly we need $r_o \geq 1$. It is desirable to have this parameter as small as possible. Indeed, in the asymptotic case ($n \rightarrow \infty$), $r_o = 1$ is achievable. This has been proved in \cite{WV09}.}.

Researchers have worked intensively in the following main areas: 1) designing \textbf{G} for given $k$ and $n$, in order to reduce the number of measurements $m$ required for a successful recovery, 2) improving the recovery algorithms for given $n$ and $m$ to be able to reconstruct signals with larger density ratio, i.e., signals with more non-zero components, and 3) analyzing performance measures of different recovery algorithms in the asymptotic regime (as $n \rightarrow \infty$) in order to compare different algorithms and be able to give an estimate of the performance for finite $n$.

Donoho in \cite{D06} and Cand\`{e}s \textit{et. al.} in \cite{CRTFeb06} used sensing matrices with i.i.d. Gaussian entries and the $\ell_1$ norm minimization of the signal estimate as the reconstruction method. Their random sensing matrices contain mostly non-zero elements, which makes the encoding computationally intense. Inspired by the good performance of sparse matrices in channel coding, some researchers (e.g. \cite{SBB206} and \cite{ZP08}) used sparse matrices, as the sensing matrix.

From the viewpoint of recovery complexity, the $\ell_1$ minimization algorithm has a computational complexity of $O(n^3)$. To reduce this complexity, some researchers used iterative algorithms as the decoder. For example, the authors in \cite{SBB206} used an iterative algorithm with a computational complexity of $O(n\cdot\log n)$ over regular bipartite graphs, while Xu and Hassibi in \cite{XH07} discussed a different iterative algorithm with a complexity of $O(n)$ based on a class of sparse graphs called \textit{expander graphs} \cite{HLW06}. Authors in \cite{DM109} and \cite{DM209} proposed and analyzed an iterative thresholding algorithm over dense graphs with complexity between $O(n\log n)$ and $O(n^2)$, depending on the sensing matrix used. The two verification-based (VB) iterative algorithms originally proposed by \cite{LM05} in the context of channel coding, were analyzed in \cite{ZP09,ZP07,ZP07J} for the case when $k/n \rightarrow 0 \text{ as } n \rightarrow \infty$ in the context of compressed sensing. In \cite{LMPDK08} and \cite{APT09} asymptotic analysis of some iterative message-passing algorithms over sparse sensing matrices can be found. The sensing matrices used in \cite{ZP09,ZP07,ZP07J,LMPDK08,APT09} are all sparse.

Our main goal in this paper is to develop a framework for the asymptotic analysis (as $n \rightarrow \infty$) of VB algorithms over sparse random sensing matrices and extend it to include recovery algorithms of similar nature such as \cite{XH07}. In our work we show that the overall computational complexity of the analysis is linear in the number of iterations. We will also show, through simulation, that VB algorithms when applied to signals with moderate lengths (in the order of $10^5$), are in good agreement with the asymptotic results. Using our approach we can perform a comprehensive study and comparison of performance/complexity trade-off of different VB recovery algorithms over a variety of sparse graphs.

The rest of the paper is organized as follows. In section \ref{Defs}, we present notations, definitions and assumptions used throughout the paper. We will also introduce VB algorithms in more detail. In section \ref{enc} the encoding process and input distributions are described. Decoding algorithms are described in section \ref{Decoding}. The analysis framework and its generalization will be introduced in sections \ref{analysis} and \ref{generalization}, respectively. Simulation results will be presented in section \ref{simulation}.
%%%%%%%%%%%%%%%%%%%%%%%%%%%%%%%%%%%%%%%%%%%%%%%%%%%%%%%%%%%%%%%%%%%%%%%%%%%%%%%%%%%%%%%%%%%%%%%%%%%%%%%%%%%%%%%%%%%
%%%%%%%%%%%%%%%%%%%%%%%%%%%%%%%%%%%%%%%%%%%%%%%%%%%%%%%%%%%%%%%%%%%%%%%%%%%%%%%%%%%%%%%%%%%%%%%%%%%%%%%%%%%%%%%%%%%
\section{Background}
\label{Defs}
%%%%%%%%%%%%%%%%%%%%%%%%%%%%%%%%%%%%%%%%%%%%%%%%%%%%%%%%%%%%%%%%%%%%%%%%%%%%%%%%%%%%%%%%%%%%%%%%%%%%%%%%%%%%%%%%%%%
\subsection{Bipartite Sensing Graph}
In general, the sensing matrix \textbf{G} can be thought as the weighted incidence matrix of a weighted bipartite graph. In this case, the element in row $i$ and column $j$ of \textbf{G} is the coefficient of the $i^\text{th}$ signal element ($v_i$) in the linear combination resulting the $j^\text{th}$ measurement $c_j$. If the weights are all 1 ($\in \mathbb{R}$), then \textbf{G} will reduce to the incidence matrix of a bipartite graph.

Consider a bipartite graph with node sets $\mathcal{V}$ and $\mathcal{C}$. Following channel coding terminology, we will call $\mathcal{V}$ the set of \textit{variable nodes} and $\mathcal{C}$ the set of \textit{check nodes}. In the compressive sensing context, signal components and measurements are mapped to variable nodes and check nodes, respectively. We will interchangeably use the terms \textit{variable nodes} and \textit{signal elements} as well as \textit{check nodes} and \textit{measurements}.

In \textit{regular} bipartite graphs, each variable node (check node) is incident to the same number $d_v$ ($d_c$) of check nodes (variable nodes). The numbers $d_v$ and $d_c$ are called \textit{variable node degree} and \textit{check node degree}, respectively. All graphs discussed in this paper are sparse regular bipartite graphs denoted by the pair $(d_v,d_c)$ and simply referred to as \textit{graphs}.

For a variable node $v_i$ we use the notation $\mathcal{M}(v_i)\subset \mathcal{C}$ to denote the set of check nodes incident to it. The graph composed of a subset $\mathcal{V}^*$ of variable nodes, their neighboring check nodes $\mathcal{M}(\mathcal{V}^*)$ and all the edges in between is called the \textit{subgraph induced by $\mathcal{V}^*$}.
%%%%%%%%%%%%%%%%%%%%%%%%%%%%%%%%%%%%%%%%%%%%%%%%%%%%%%%%%%%%%%%%%%%%%%%%%%%%%%%%%%%%%%%%%%%%%%%%%%%%%%%%%%%%%%%%%%%
\subsection{Verification Based Algorithms}
\label{VB}
Two iterative algorithms over bipartite graphs are proposed and analyzed in \cite{LM05} for packet-based error correction in the context of channel coding. In these algorithms, a variable node can be in one of the two states: ``verified'' or ``unknown''. Under certain circumstances, a variable node is verified and a value is assigned to it. This variable node, then contributes to the verification of other nodes. The decoding process continues until either the unknown variable nodes become verified entirely, or the process makes no further progress. Due to the verification nature of the procedure, the two algorithms in \cite{LM05} are called \textit{verification-based} (VB) algorithms. When used in the context of compressive sensing, we would like to see VB algorithms to correctly verify signal elements in each iteration. Indeed, in section \ref{enc}, we define sufficient conditions for VB algorithms to result in the original signal.

As noted in \cite{ZP07J}, authors in \cite{LM05} defined the two VB algorithms using node-based (NB) representation but analyzed them using message-based (MB) representation. In the NB representation, the ``verified'' state of a variable node is a property of the node itself. In the MB representation, however, the state is reflected in the outgoing messages from a variable node. Therefore, in contrast to the NB case, multiple different states may exist for the same variable node. In \cite{ZP07J}, authors showed that for one of the algorithms, the two versions perform the same. But for the other algorithm, the NB version outperforms the MB one (in compressive sensing this implies that NB version can successfully recover signals with larger density ratios).

A well-known method to analyze such iterative algorithms in coding theory is density evolution \cite{RU01}. However, as density evolution can only be applied to the MB representations, authors in \cite{ZP07J} used differential equations to analyze the NB versions in the case where $n \rightarrow \infty$. Applying their analysis to $(d_v,d_c)$ graphs, the number of differential equations is roughly $(d^2_c+3d_c)/2$, which becomes intractable for large $d_c$. Therefore, authors used numerical calculations to see the success/failure of the NB algorithms.

In the context of compressive sensing, authors in \cite{ZP09} analyzed the MB-VB algorithms using density evolution for super-sparse signals ($k/n \rightarrow 0 \text{ as } n \rightarrow \infty$). In our work, we analyze NB-VB algorithms in the regime where $n \rightarrow \infty$ and $k$ grows linearly with $n$. In section \ref{analysis} we show that the complexity of our methodology is less than the one used in \cite{ZP07J}.
%%%%%%%%%%%%%%%%%%%%%%%%%%%%%%%%%%%%%%%%%%%%%%%%%%%%%%%%%%%%%%%%%%%%%%%%%%%%%%%%%%%%%%%%%%%%%%%%%%%%%%%%%%%%%%%%%%%
%%%%%%%%%%%%%%%%%%%%%%%%%%%%%%%%%%%%%%%%%%%%%%%%%%%%%%%%%%%%%%%%%%%%%%%%%%%%%%%%%%%%%%%%%%%%%%%%%%%%%%%%%%%%%%%%%%%
\section{Encoding and Input Distribution}
\label{enc}
Let $\mathcal{K}$ denote the set of non-zero elements in the original signal. We refer to this set as the \textit{support set}. In general, there are two ways to define signal elements in compressive sensing:
\begin{enumerate}
	\item Let $k = |\mathcal{K}| = \gamma n$ be a deterministic value. Out of $n$ signal elements, $k$ of them are selected at random as the support set. The value of each non-zero element is then an i.i.d. random variable with probability distribution $f$.
	\item Let $\alpha$, referred to as \textit{density factor}, be the probability that a signal element belongs to $\mathcal{K}$. By fixing $\alpha$, each of the $n$ signal elements has a value i.i.d. with the following distribution: it is zero with probability $1-\alpha$ or follows a distribution $f$ with probability $\alpha$. In this case, $k$ and $\gamma \triangleq k/n$ are random variables. Furthermore, $E[\gamma] = \alpha$ and $E[k] = \alpha n$, where $E[\cdot]$ denotes the expected value.
\end{enumerate}
When $n \rightarrow \infty$, as a consequence of law of large numbers, both cases (1) and (2) provide the same results. In the rest of the paper, we adopt the second model. In this paper, we show that using NB-VB recovery algorithms in the asymptotic regime as $n \rightarrow \infty$, a limiting value exists for $\alpha$, before which the recovery algorithm is successful and beyond which it is not. Henceforth, we refer to this limit as \textit{success threshold}.

The weights of the bipartite graph, corresponding to the non-zero entries of the sensing matrix \textbf{G}, can be drawn i.i.d. according to a distribution $g$. In this work, we make the assumption that at least one of the distributions $f$ or $g$ is continuous. Similar conditions have been used in \cite{ZP08} and \cite{SBB206}. As a consequence, we introduce and prove Theorem \ref{unique} below. 
\begin{theorem}
\label{unique}
Let $c_i$ and $c_j$ be two distinct check nodes and $\mathcal{V}_i$ and $\mathcal{V}_j$ be their corresponding set of incident variable nodes in $\mathcal{K}$; i.e., $\mathcal{V}_i=\mathcal{M}(c_i)\cap \mathcal{K}$ and $\mathcal{V}_j=\mathcal{M}(c_j)\cap \mathcal{K}$. Suppose at least one of the distributions $f$ or $g$ described before is continuous. If $c_i = 0$ then $\mathcal{V}_i$ is the empty set with probability one. Moreover if $\mathcal{V}_i\neq \mathcal{V}_j$ then:
\[
\Pr\left(c_i = c_j\right) = 0.
\]
\end{theorem}
Before proving the theorem, let's state the \textit{Uniqueness of Samples} fact, which is used in the proof.
\begin{fact}[Uniqueness of Samples]
Let $x_i$ and $x_j$ be two independent samples drawn from a continuous distribution. It follows that:
\[
\Pr\left(x_i = x_j\right) = 0.
\]
In other words, no two independent continuous samples will have the same value, almost surely. More generally, if $c$ denotes any constant, then
\[
\Pr\left(x_i = c\right) = 0.
\]
\end{fact}
\begin{IEEEproof}[Proof of Theorem \ref{unique}]
The value of a check node $c_j$ is $\sum_{i:v_i\in\mathcal{M}(c_j)}{w_{ij}v_i}$, where $w_{ij}$ is the weight associated with the edge connecting the variable node $v_i$ to check node $c_j$. Thus, a check node will have a continuous distribution whenever at least one of its neighboring variable nodes belongs to the support set, and will be zero otherwise. The proof is then complete according to the \textit{Uniqueness of Samples} fact.
\end{IEEEproof}
Based on Theorem \ref{unique} the following statements are correct with probability one (almost surely):
\begin{itemize}
	\item[S1:] if two check nodes $c_i$ and $c_j$ have the same non-zero value, they are both neighbor to the same elements of the set $\mathcal{K}$, i.e., $\{\mathcal{M}(c_i) \cap \mathcal{K}\} \equiv \{\mathcal{M}(c_j) \cap \mathcal{K}\}$.
	\item[S2:] if the value of a check node $c_i$ is zero, none of its neighboring variable nodes belongs to the set $\mathcal{K}$, i.e., $\{\mathcal{M}(c_i) \cap \mathcal{K}\} \equiv \emptyset$.
\end{itemize}
In VB algorithms, as we will see in the next section, variable nodes are verified based on similar statements as S1 and S2. Therefore, the assumption on the continuity of $f$ or $g$ is a sufficient condition for the algorithms to converge to the true original signal. Henceforth, we assume that all the weights of the bipartite graph (and therefore the entries of \textbf{G}) are in $\{0,1\}$, and the distribution $f$ is continuous over $\mathbb{R}$. 
%%%%%%%%%%%%%%%%%%%%%%%%%%%%%%%%%%%%%%%%%%%%%%%%%%%%%%%%%%%%%%%%%%%%%%%%%%%%%%%%%%%%%%%%%%%%%%%%%%%%%%%%%%%%%%%%%%%
%%%%%%%%%%%%%%%%%%%%%%%%%%%%%%%%%%%%%%%%%%%%%%%%%%%%%%%%%%%%%%%%%%%%%%%%%%%%%%%%%%%%%%%%%%%%%%%%%%%%%%%%%%%%%%%%%%%
\section{Decoding Process and Recovery Algorithms}
\label{Decoding}
The decoder, knowing the measurements and the sensing matrix, tries to recover the original signal. So, neither the density factor $\alpha$ nor the support set is known at the decoder.

In this section we discuss the first algorithm (LM1) used in \cite{ZP09} (here referred to as LM) and the algorithm used in \cite{SBB206} (referred to as SBB). These two algorithms are the original VB algorithms in the context of compressive sensing. With the description given in section \ref{VB}, the algorithm in \cite{XH07}, referred to as XH, falls in the category of VB algorithms as well. In the original XH, at each iteration, only one variable node is verified. Here we propose and discuss a parallel version of this algorithm. For the case where $n \rightarrow \infty$ and $d_v \geq 5$, analyzing the set of all variable nodes that can be verified potentially at each iteration of the original XH, it can be shown that the verification of one variable node does not exclude another variable node from this set. Therefore, both versions of XH perform identically in terms of success threshold. The parallel version, however, is considerably faster.

As the last algorithm, we reveal the support set to the decoder and use the conventional peeling decoder in \cite{LMSS01}. We will refer to this algorithm as \textit{Genie}. The Genie performance will be an upper bound on the performance of VB algorithms\footnote{The performance of Genie is the same as the performance of peeling algorithm over BEC.}.

The description of these four algorithms follows next. Except for the Genie, all variable nodes are initially ``unknown''. Before the first iteration, all variable nodes that have at least one neighboring check node with the value equal to 0, are removed from the graph. In each iteration of the four algorithms, when a variable node is verified, its verified value is subtracted from the value of all neighboring check node. The node is then removed from the graph along with all edges adjacent to it. Check nodes with degree 0 are also removed from the graph. At any iteration, the algorithms stop if either all variable nodes are verified, or the algorithm makes no further progress.

At each iteration $\ell$, the algorithms proceed as follows:

\textbf{LM}
\begin{itemize}
	\item  find the degree of each check node. Verify all variable nodes that have at least one check node of degree one with the value of singly-connected check nodes.
\end{itemize}

\textbf{SBB}
\begin{enumerate}
	\item sequentially go through all variable nodes. For each variable node $v$, look for two check nodes $c_i, c_j \in \mathcal{M}(v), j\neq i$ with identical value $g$.
	\item Verify all variable nodes $v'$ adjacent to either $c_i$ or $c_j$, i.e., $\forall v': v' \in \{\mathcal{M}(c_i)\cup\mathcal{M}(c_j)\} - \{\mathcal{M}(c_i)\cap\mathcal{M}(c_j)\}$, to zero.
	\item If $v$ is the only variable node connected to both $c_i$ and $c_j$ ($v \equiv \{\mathcal{M}(c_i)\cap\mathcal{M}(c_j)\}$) verify it with the value $g$.
\end{enumerate}
For the sake of presentation we have presented simplified versions of LM and SBB algorithms. In section \ref{generalization}, the necessary modifications that should be made in the analysis to deal with the original algorithms are discussed.

\textbf{XH}
\begin{itemize}
	\item find all variable nodes $v_i$ such that for each $v_i$, $\lceil d_v/2 \rceil$ or more of its neighboring check nodes ($\mathcal{M}(v_i)$) have the same value $g_i$. For each such variable node, verify $v_i$ by the value $g_i$.
\end{itemize}

\textbf{Genie Algorithm}
\begin{itemize}
	\item in the subgraph induced by the unverified variable nodes in the set $\mathcal{K}$, verify all variable nodes that have at least one check node of degree one with the value of singly-connected check nodes.
\end{itemize}
%%%%%%%%%%%%%%%%%%%%%%%%%%%%%%%%%%%%%%%%%%%%%%%%%%%%%%%%%%%%%%%%%%%%%%%%%%%%%%%%%%%%%%%%%%%%%%%%%%%%%%%%%%%%%%%%%%%
%%%%%%%%%%%%%%%%%%%%%%%%%%%%%%%%%%%%%%%%%%%%%%%%%%%%%%%%%%%%%%%%%%%%%%%%%%%%%%%%%%%%%%%%%%%%%%%%%%%%%%%%%%%%%%%%%%%
\section{Asymptotic Analysis Framework}
\label{analysis}
\begin{figure}[h]
\begin{minipage}[b]{0.5\linewidth}
\centering
\includegraphics[height=100 pt]{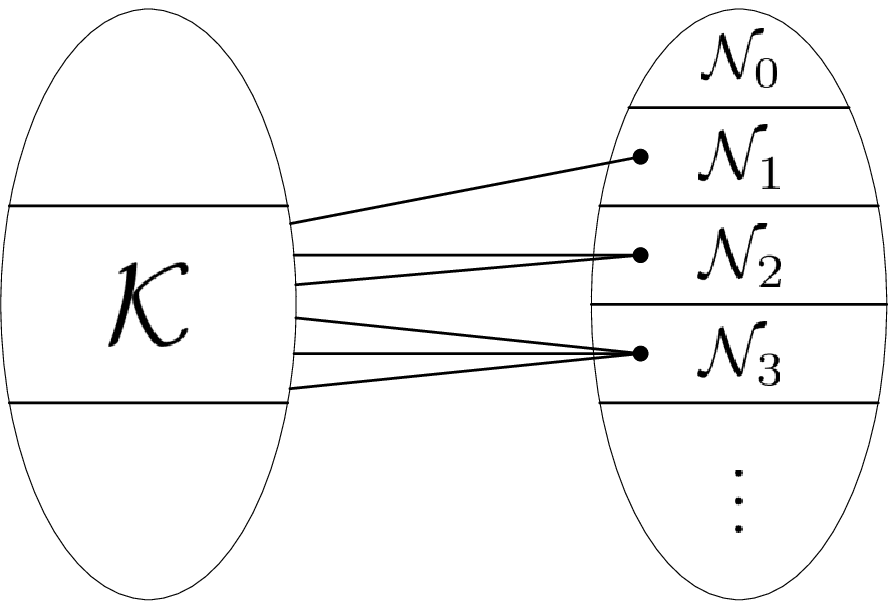}
\caption{Partitioning check nodes based on their degree in the\newline subgraph induced by $\mathcal{K}$.}
\label{N1}
\end{minipage}
%\hspace{0.5cm}
\begin{minipage}[b]{0.5\linewidth}
\centering
\includegraphics[height=100 pt]{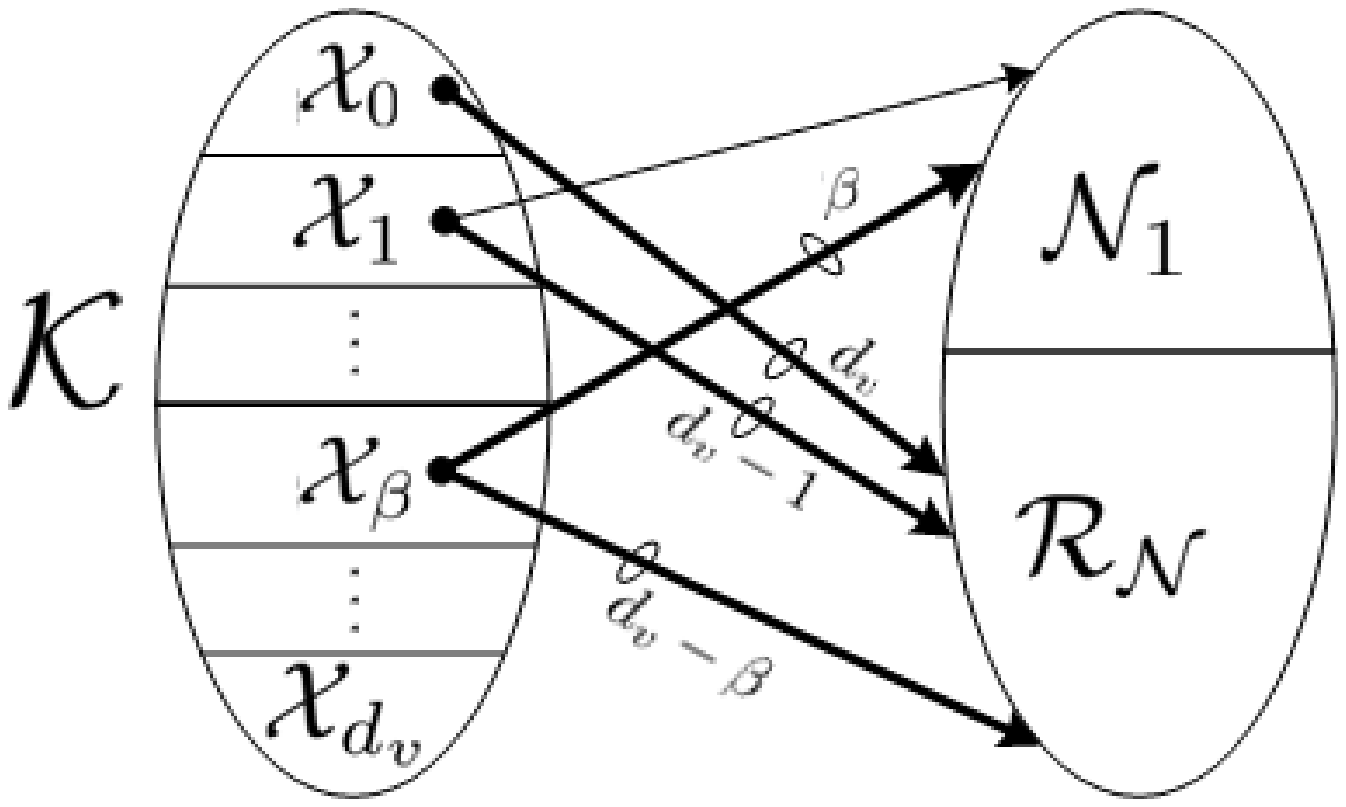}
\caption{Partitioning variable nodes in $\mathcal{K}$ based on the number of their neighbors in $\mathcal{N}_1$.}
\label{K1}
\end{minipage}
\end{figure}

To describe the analysis framework, we assume a ($d_v,d_c$) graph. Let $\mathcal{V}^* \subset \mathcal{V}$ be a subset of variable nodes and $\mathcal{G}^*$ be the left-regular graph induced by the set $\mathcal{V}^*$. We denote the set of check nodes with degree $i,1\leq i\leq d_c$ in $\mathcal{G}^*$ by $\mathcal{N}_i$. This partitioning is depicted in Figure \ref{N1} for $\mathcal{V}^* \equiv \mathcal{K}$. For mathematical convenience we let $\mathcal{N}_0$ be the set of check nodes that have been removed from the induced subgraph. Clearly, $\mathcal{C} = \bigcup_{i=0}^{d_c}\mathcal{N}_i$. Further, let $\mathcal{X}_i \subset \mathcal{V}^*,1\leq i\leq d_v$ be the set of variable nodes that have $i$ edges connected to the set $\mathcal{N}_1$. Figure \ref{K1} shows the partitioning of $\mathcal{V}^* \equiv \mathcal{K}$ to $\mathcal{X}_i$s.

At this stage, we model the verification process of Genie, LM, SBB and XH algorithms using the sets $\mathcal{X}_i$'s in the asymptotic regime where $n \rightarrow \infty$. This verification model is presented and proved in Theorem \ref{Model_Iter}. In this theorem, $\mathcal{K}' \triangleq \mathcal{K}\cup\mathcal{K}_\Delta$, where $\mathcal{K}_\Delta$ is the set of zero-valued variable nodes, in which all variable nodes have $d_v$ edges connected to the set $\mathcal{N}_S \triangleq \bigcup_{i=1}^{d_c} \mathcal{N}_i$.
\begin{theorem}
\label{Model_Iter}
In each iteration, a variable node is verified asymptotically almost surely if and only if it belongs to the set $\bigcup_{i=\beta}^{d_v}\mathcal{X}_i$, where $\beta$ equals 1, 2, $\lceil d_v/2 \rceil$ for the Genie, SBB and XH, respectively. In these cases $\mathcal{V}^* \equiv \mathcal{K}$.\\
In each iteration of the LM algorithm a variable node is verified asymptotically almost surely if and only if it belongs to the set $\bigcup_{i=1}^{d_v}\mathcal{X}_i$. In this case $\mathcal{V}^* \equiv \mathcal{K}'$.
\end{theorem}
\begin{IEEEproof}[Proof of Theorem \ref{Model_Iter}]
We first prove the theorem for the SBB algorithm. The proof can be used also for the XH algorithm with no major changes. A variable node $v$ is resolved in the SBB algorithm if and only if it is the only unresolved variable node attached to at least two check nodes $c_{i_1}$ and $c_{i_2}$ with the same value. If $c_{i_1},c_{i_2}\in\mathcal{N}_1$, then by definition $v\in\left\{\mathcal{X}_2,\mathcal{X}_3,\cdots,\mathcal{X}_{d_v}\right\}$.\\
To prove the converse, assume that at iteration $\ell$, $v\in\mathcal{X}_2$ for simplicity. The only way that this variable node is not resolved in this iteration is that it shares the two singly-connected check nodes with at least one zero-valued variable node $v'$. However, this means that the two variable nodes $v,v'$ form a cycle of length 4. According to \cite{MWW04}, a random regular graph has a fixed number of short cycles regardless of its size. Thus, tending the number of variable nodes to infinity, the probability that two variable nodes $v,v'$ form a cycle of length 4 goes to zero. In other words, variable node $v\in\left\{\mathcal{X}_2,\mathcal{X}_3,\cdots,\mathcal{X}_{d_v}\right\}$ is resolved in the SBB algorithm asymptotically almost surely. This completes the proof. 

In the LM algorithm, after removing variable nodes with at least one check node with the value equal to zero, the remaining variable nodes in the graph are in the set $\mathcal{K}'$. This justifies the use of $\mathcal{K}'$. The rest of the proof follows as before.
\end{IEEEproof}
\begin{corollary}
\label{cor}
For a ($d_v,d_c$) graph, the success threshold is the highest for Genie, followed by SBB and lastly XH. This is because the number of $\mathcal{X}_i$ sets contributing to the verification of variable nodes decreases in the same order.
\end{corollary}
Corollary \ref{cor} is also verified in the simulation section \ref{simulation}. Theorem \ref{Model_Iter} allows us to model the sensing graph and its evolution in the four algorithms with a graph induced by the support set $\mathcal{K}$ or $\mathcal{K}'$ along with the evolution of the sets $\mathcal{X}_i$ and $\mathcal{N}_j$ in the asymptotic regime. To formulate the evolution of the sets $\mathcal{N}_i$ and $\mathcal{X}_i$, we denote by $\raisebox{2pt}{$p$}^{(\ell)}_{\mathcal{N}_i}$ ($\raisebox{2pt}{$p$}^{(\ell)}_{\mathcal{X}_i}$), the set of probabilities that a check node (variable node) belongs to the set $\mathcal{N}_i$ ($\mathcal{X}_i$) at iteration $\ell$. The superscript $(\ell)$ denotes the iteration number $\ell$. Also, we denote by $\alpha^{(\ell)}$ the probability that a variable node belongs to the unverified set $\mathcal{K}^{(\ell)}$. An iteration $\ell \geq 1$ starts by knowing the probabilities $\alpha^{(\ell)}$, $\raisebox{2pt}{$p$}^{(\ell)}_{\mathcal{N}_i}$ and $\raisebox{2pt}{$p$}^{(\ell)}_{\mathcal{X}_i}$, continues by the calculation of $\alpha^{(\ell+1)}$, and ends with the calculation of $\raisebox{2pt}{$p$}^{(\ell+1)}_{\mathcal{N}_i}$ and $\raisebox{2pt}{$p$}^{(\ell+1)}_{\mathcal{X}_i}$. 

Using this analysis approach we are able to track the evolution of $\alpha^{(\ell)}$ with iteration for a given initial density factor $\alpha^{(0)}$. The analysis proceeds until either the probability $\alpha^{(\ell)}$ decreases monotonically to zero as the number of iterations increase, or it is bounded away from zero for any number of iterations. In the first case the algorithm succeeds in recovering the original signal entirely, while in the second case it fails. By examining different values of $\alpha^{(0)}$, the success threshold, defined as the supremum value of $\alpha^{(0)}$ for which the signal can be fully recovered as $n\rightarrow\infty$ and $\ell\rightarrow\infty$, can be determined for different $(d_v,d_c)$ pairs.

In what follows, we show the algorithm to find the update rules for different probabilities. The formulas are calculated using combinatorial enumeration and probabilistic arguments. The proofs can be found in Appendix \ref{Details}.
\begin{enumerate}
	\item Based on the set of probabilities $\raisebox{2pt}{$p$}^{(\ell)}_{\mathcal{X}_i}$, find the probability $\raisebox{2pt}{$p$}^{(\ell)}_{r}$, that a variable node is verified in iteration $\ell+1$ from $\raisebox{2pt}{$p$}^{(\ell)}_{r} = \sum_{i=\beta}^{d_v}{\raisebox{2pt}{$p$}^{(\ell)}_{\mathcal{X}_i}}$. The value of $\beta$ can be 1, or 2, or $\lceil d_v/2 \rceil$ as in Theorem \ref{Model_Iter}. The probability $\alpha^{(\ell+1)}$ then follows from $\alpha^{(\ell+1)} = \alpha^{(\ell)}(1-\raisebox{2pt}{$p$}^{(\ell)}_r)$.
	\item Find the set of probabilities $\raisebox{2pt}{$p$}^{(\ell+1)}_{\mathcal{N}_j},j=0,\cdots,d_c$ from $\raisebox{2pt}{$p$}^{(\ell+1)}_{\mathcal{N}_j} = \sum_{i=j}^{d_c}{p_{\mathcal{N}_{i}}^{(\ell)}p_{\mathcal{N}_{ij}}^{(\ell)}}$, where:
\[
\raisebox{2pt}{$p$}^{(\ell)}_{\mathcal{N}_{10}} = \displaystyle\frac{\alpha^{(\ell)}d_c\displaystyle\sum_{i=\beta}^{d_v}{i\raisebox{2pt}{$p$}^{(\ell)}_{\mathcal{X}_i}}}{d_v \raisebox{2pt}{$p$}^{(\ell)}_{\mathcal{N}_1}},\hspace{1cm}\raisebox{2pt}{$p$}^{(\ell)}_{\mathcal{N}_{11}} = 1-\raisebox{2pt}{$p$}^{(\ell)}_{\mathcal{N}_{10}}, \hspace{1cm}
\raisebox{2pt}{$p$}^{(\ell)}_{\mathcal{N}_{ij}} = {i\choose{i-j}} \left(A\right)^{i-j}\left(1-A\right)^j,\hspace{10pt} i\geq2.
\]
	\item Find the set of probabilities $\raisebox{2pt}{$p$}^{(\ell+1)}_{\mathcal{X}_i},i=0,\cdots,d_v$ from $\raisebox{2pt}{$p$}^{(\ell+1)}_{\mathcal{X}_i} = \frac{1}{N}\sum_{j=0}^{\min\{i,\beta-1\}}{p_{\mathcal{X}_j}^{(\ell)}p_{\mathcal{X}_{ji}}^{(\ell)}}$, where:
\[
\raisebox{2pt}{$p$}_{\mathcal{X}_{ij}}^{(\ell)} = {d_v-i\choose j-i}\left(B\right)^{j-i}\left(1-B\right)^{d_v-j}, \hspace{10pt} N = 1-p_r^{(\ell)}.
\]
The quantities $A$ and $B$ are given by:
\[
\hspace{-10pt}
A = 
\frac{d_v \raisebox{2pt}{$p$}^{(\ell)}_{r} - \displaystyle\sum_{i=\beta}^{d_v}{i\raisebox{2pt}{$p$}^{(\ell)}_{\mathcal{X}_i}}}{d_v\left(1 - \frac{\displaystyle\raisebox{2pt}{$p$}^{(\ell)}_{\mathcal{N}_1}}{\displaystyle\alpha^{(\ell)} d_c}\right)},
\hspace{10pt}
B = \frac{\displaystyle\sum_{j=2}^{d_c}{\raisebox{2pt}{$p$}_{\mathcal{N}_j}^{(\ell)}\raisebox{2pt}{$p$}_{\mathcal{N}_{j1}}^{(\ell)}}}{\displaystyle\sum_{j=2}^{d_c}{\raisebox{2pt}{$p$}_{\mathcal{N}_j}^{(\ell)}\raisebox{2pt}{$p$}_{\mathcal{N}_{j1}}^{(\ell)}}+\displaystyle\sum_{i=2}^{d_c}{i\raisebox{2pt}{$p$}_{\mathcal{N}_i}^{(\ell+1)}}}.
\]
\end{enumerate}

The initial probabilities $\raisebox{2pt}{$p$}^{(0)}_{\mathcal{N}_i}$ and $\raisebox{2pt}{$p$}^{(0)}_{\mathcal{X}_i}$ for Genie, SBB and XH are as follows. These probabilities for the LM are rather more involved and can be seen in Appendix \ref{Details}.
\[
\raisebox{2pt}{$p$}^{(0)}_{\mathcal{N}_i} = {d_c\choose i} \left(\alpha^{(0)}\right)^i \left(1-\alpha^{(0)}\right)^{d_c-i},\hspace{20pt}i=0,\cdots,d_c.
\]
\[
\raisebox{2pt}{$p$}^{(0)}_{\mathcal{X}_i} = {d_v\choose i}\left(p^{{(0)}}\right)^i\left(1-p^{(0)}\right)^{d_v-i},\hspace{20pt}i=0,\cdots,d_v,
\]
where, $p^{{(0)}} = \raisebox{2pt}{$p$}^{(0)}_{\mathcal{N}_1}/\alpha^{(0)}d_c$, and $\alpha^{(0)} = \alpha$ (density factor). The number of update rules in each iteration, is almost the same as the one in \cite{ZP07J}. Therefore, the overall computational complexity of the analysis is linear in $\ell$. However, the calculation of the updates is based on simple calculations as opposed to solving differential equations as in \cite{ZP07J}.
%%%%%%%%%%%%%%%%%%%%%%%%%%%%%%%%%%%%%%%%%%%%%%%%%%%%%%%%%%%%%%%%%%%%%%%%%%%%%%%%%%%%%%%%%%%%%%%%%%%%%%%%%%%%%%%%%%%
%%%%%%%%%%%%%%%%%%%%%%%%%%%%%%%%%%%%%%%%%%%%%%%%%%%%%%%%%%%%%%%%%%%%%%%%%%%%%%%%%%%%%%%%%%%%%%%%%%%%%%%%%%%%%%%%%%%
\section{Generalization of the Framework}
\label{generalization}
The extra step in the original LM and SBB algorithms is as follows: at each iteration, look for check nodes with value equal to zero. If such a check node is found, verify the neighboring variable nodes with the value zero. 

To analyze the original algorithms, the set of variable nodes is divided in two sets $\mathcal{K}$ and $\mathcal{K}'$ as in the simplified LM. The set of check nodes is then categorized into sets $\mathcal{N}_{ij}$, where $i$ and $j$ indicate the number of edges between the check node and the sets $\mathcal{K}$ and $\mathcal{K}'$, respectively. The recursive formulas for the new setup can be found using the same methodology as before.
%%%%%%%%%%%%%%%%%%%%%%%%%%%%%%%%%%%%%%%%%%%%%%%%%%%%%%%%%%%%%%%%%%%%%%%%%%%%%%%%%%%%%%%%%%%%%%%%%%%%%%%%%%%%%%%%%%%
%%%%%%%%%%%%%%%%%%%%%%%%%%%%%%%%%%%%%%%%%%%%%%%%%%%%%%%%%%%%%%%%%%%%%%%%%%%%%%%%%%%%%%%%%%%%%%%%%%%%%%%%%%%%%%%%%%%
\section{Simulation Results}
\label{simulation}
In this section, we present simulation results obtained by running the recovery algorithms over random regular bipartite graphs to recover sparse signals of finite length $n$. We also present analytical results obtained by running the mathematical analysis described in Section \ref{analysis} for the asymptotic regime of $n\rightarrow\infty$. The comparison of the results shows that there is a good agreement between simulation and analytical results for moderately large $n = 10^5$.

In all simulations, signal elements (variables) are drawn from a Gaussian distribution. The regular bipartite graphs are constructed randomly with no parallel edges and all the edge weights equal to one. Each simulation point is generated by averaging over 1000 random instances of the input signal.

For the analytical results, we consider the recovery algorithm successful if $\alpha^{(\ell)} < 10^{-7}$. To calculate the success threshold, a binary search is performed until the separation between start and end of the search region is less than $10^{-4}$.

\begin{figure}[!h]
\centering
\includegraphics[height=280 pt]{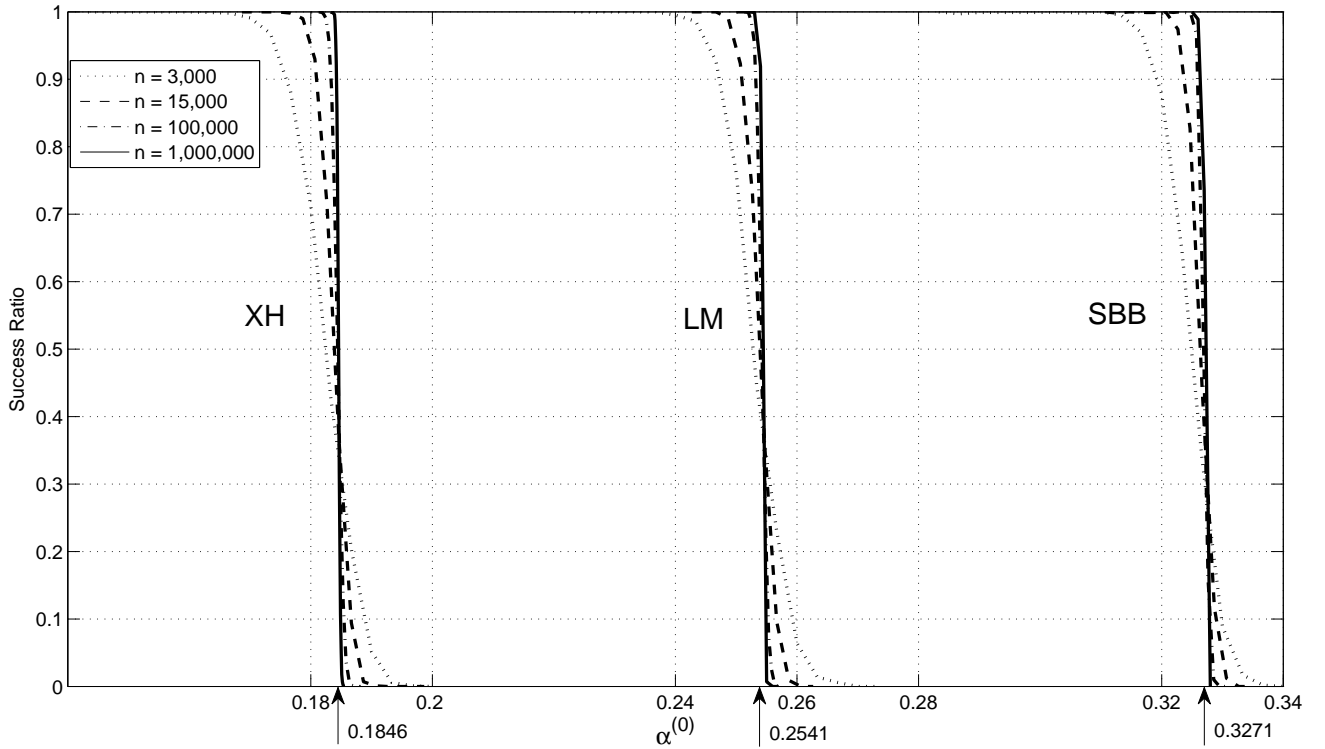}
\caption{Success Ratio of Algorithms XH, LM and SBB vs. $\alpha^{(0)}$ for $(5,6)$ graphs with $n=3K, 15K, 100K \text{ and } 1000K$. Analytical thresholds are shown by arrows.}
\label{VarLength}
\end{figure}

As the first experiment, we apply XH, SBB and LM algorithms to four randomly constructed (5,6) regular graphs with $n=\{3, 15, 100, 1000\}\times 10^3$. The success rate of the algorithms vs. the initial density factor $\alpha^{(0)}$ are shown in Figure \ref{VarLength}. From the figure, we can see that, for all algorithms, by increasing $n$, the transition part of the curves becomes sharper such that the curves for $n=10^6$ practically look like a step function. In the figure we have also given the success threshold of the algorithms for $(5,6)$ graphs, obtained based on the proposed analysis, as arrows. As can be seen, the thresholds match very well with the waterfall region of the simulation curves.

In Table \ref{success_threshold_1}, we have listed the analytical success thresholds of the iterative recovery algorithms for graphs with different $d_v$ and $d_c$. The results for XH and SBB algorithms on $(3,4)$ graphs are missing as these algorithms perform poorly on such graphs. As expected, for every graph, the Genie algorithm has the best performance. This is followed by SBB, LM and XH algorithms, respectively. Careful inspection of the results in Table \ref{success_threshold_1} indicates that the oversampling ratio $r_o$, improves consistently by decreasing both $d_v$ and $d_c$ values. In fact, among the results presented in Table \ref{success_threshold_1}, the application of the Genie and LM to $(3,4)$ graphs results in the lowest oversampling ratio ($r_o = d_v/\alpha d_c$) of $\approx 1.16$ and $\approx 2.51$, respectively. Note that the success threshold of the Genie over regular graphs is far from the optimal achievable success threshold $d_v/d_c$ proved in \cite{WV09}.

\begin{figure}[t]
\centering
\includegraphics[height=230 pt]{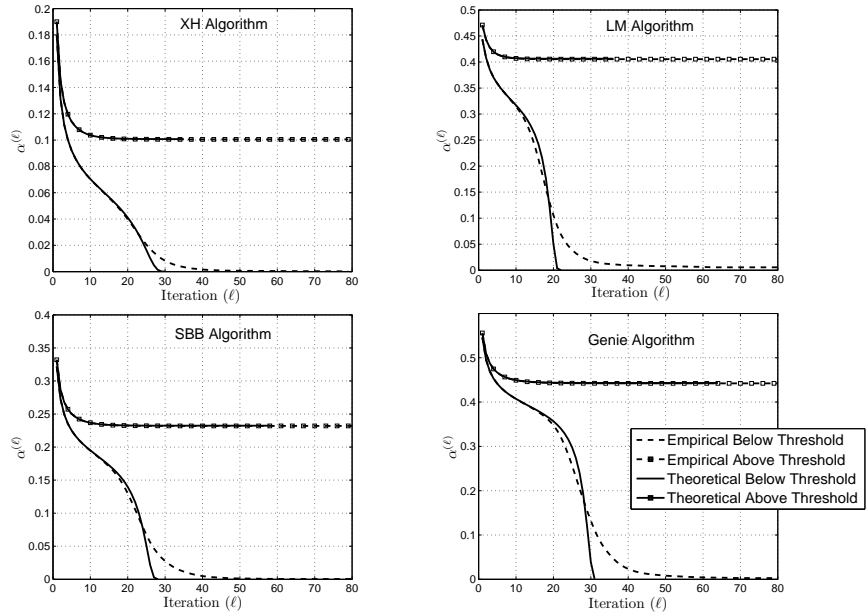}
\caption{Evolution of $\alpha^{(\ell)}$ vs. iteration number $\ell$ for the four recovery algorithms over a $(5,6)$ graph with $n=100K$.}
\label{G4G6Evolution100k}
\end{figure}

To further investigate the degree of agreement between our asymptotic theoretical analysis and finite-length simulation results, we have presented in Figure \ref{G4G6Evolution100k} the evolution of density factor $\alpha^{(\ell)}$ with iterations $\ell$ for the four algorithms over a $(5,6)$ graph with $n=10^5$. For each algorithm, two values of $\alpha^{(0)}$ are selected: one above and one below the success threshold presented in Table \ref{success_threshold_1}. The theoretical results are shown by solid lines while simulations are presented with dotted lines. As one can see, the two sets of results are in close agreement particularly for the cases where $\alpha^{(0)}$ is above the threshold and for smaller values of $\ell$.
%\setcounter{figure}{0}
%\appendices
%%%%%%%%%%%%%%%%%%%%%%%%%%%%%%%%%%%%%%%%%%%%%%%%%%%%%%%%%%%%%%%%%%%%%%%%%%%%%%%%%%%%%%%%%%%%%%%%%%%%%%%%%%%%%%%%%%%
%\section{Table and Figures}
%\label{plots}
\vspace{3cm}
\begin{table}[!h]
	\caption{Success Thresholds for different graphs and algorithms}
	\centering	
%	{\footnotesize{
	\begin{tabular}{|l|c|c|c|c|c|}
		\hline
		$(d_v,d_c)$ & $(3,4)$ & $(5,6)$ & $(5,7)$ & $(5,8)$ & $(7,8)$\\
		\hline
		\hline
		 XH& - & 0.1846 & 0.1552 & 0.1339 & 0.1435\\
		\hline
		 SBB& - & 0.3271 & 0.2783 & 0.2421 & 0.3057\\
		\hline
		 LM & 0.2993 & 0.2541 & 0.2011 & 0.1646 & 0.2127\\
		\hline
		 Genie& 0.6474 & 0.5509 & 0.4786 & 0.4224 & 0.4708\\
		\hline
	\end{tabular}
%	}}
	\label{success_threshold_1}
\end{table}
%%%%%%%%%%%%%%%%%%%%%%%%%%%%%%%%%%%%%%%%%%%%%%%%%%%%%%%%%%%%%%%%%%%%%%%%%%%%%%%%%%%%%%%%%%%%%%%%%%%%%%%%%%%%%%%%%%%
%%%%%%%%%%%%%%%%%%%%%%%%%%%%%%%%%%%%%%%%%%%%%%%%%%%%%%%%%%%%%%%%%%%%%%%%%%%%%%%%%%%%%%%%%%%%%%%%%%%%%%%%%%%%%%%%%%%
\bibliographystyle{IEEEtran}
\bibliography{arXiv}
%%%%%%%%%%%%%%%%%%%%%%%%%%%%%%%%%%%%%%%%%%%%%%%%%%%%%%%%%%%%%%%%%%%%%%%%%%%%%%%%%%%%%%%%%%%%%%%%%%%%%%%%%%%%%%%%%%%
%%%%%%%%%%%%%%%%%%%%%%%%%%%%%%%%%%%%%%%%%%%%%%%%%%%%%%%%%%%%%%%%%%%%%%%%%%%%%%%%%%%%%%%%%%%%%%%%%%%%%%%%%%%%%%%%%%%
\newpage
\appendices
\section{Detailed Description of the Analysis Framework}
\label{Details}
\textit{\textbf{B.1. General Setup}}\\
To derive the formulation for the general framework, we assume that we are at iteration $\ell$. The state of the system at this iteration is fully characterized by the set $\mathcal{K}^{(\ell)}$ and probabilities $\raisebox{2pt}{$p$}^{(\ell)}_{\mathcal{X}_i}$,  $\raisebox{2pt}{$p$}^{(\ell)}_{\mathcal{N}_i}$, and $\alpha^{(\ell)}$.\\
The probabilities $\raisebox{2pt}{$p$}^{(\ell)}_{\mathcal{N}_i}$ denote the probability of a check node having $i$ connected edges to the set $\mathcal{K}^{(\ell)}$.\\
The probabilities $\raisebox{2pt}{$p$}^{(\ell)}_{\mathcal{X}_i}$ denote the probability of a variable node in the set $\mathcal{K}^{(\ell)}$ having $i$ connected edges to the set $\mathcal{N}_1^{(\ell)}$.\\
The probability $\alpha^{(\ell)}$ denotes the probability of a variable node belonging to the set $\mathcal{K}^{(\ell)}$.\\
Throughout the analysis, the head and tail of an edge $e$ will be denoted by $h_e$ and $t_e$, respectively. As the direction of edges is of no consequence to our analysis, without loss of generality, we assign the head to the variable side and the tail to the check side.\\

\begin{figure}[ht]
\begin{minipage}[b]{0.5\linewidth}
\centering
\includegraphics[height=100 pt]{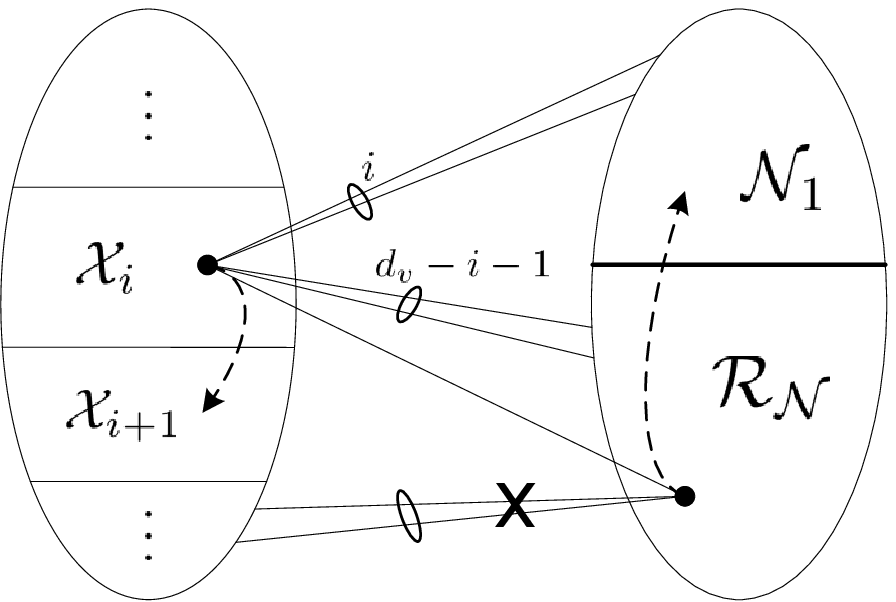}
\caption{A variable node in $\mathcal{X}_i$ turns to a variable node in $\mathcal{X}_{i+1}$ when \newline all but one of the edges connected to its adjacent check node are \newline removed from the induced subgraph.}
\label{rec}
\end{minipage}
%\hspace{0.5cm}
\begin{minipage}[b]{0.5\linewidth}
\centering
\includegraphics[height=80 pt]{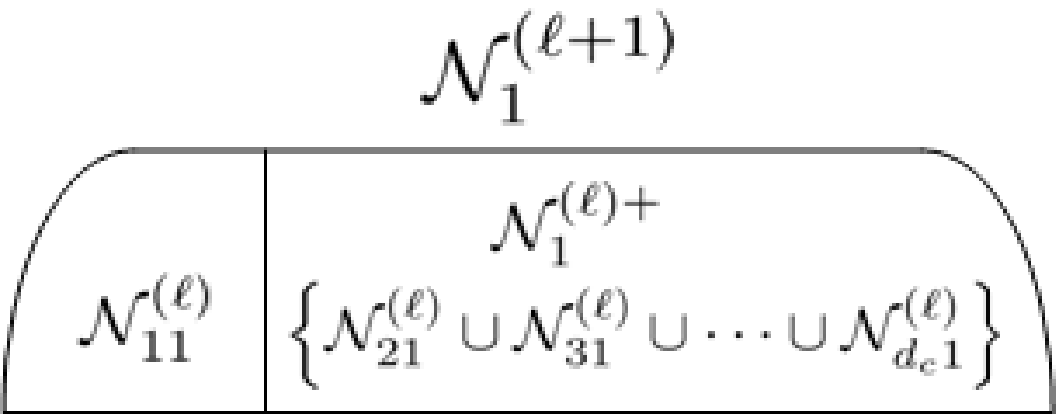}
\caption{Configuration of $\mathcal{N}_1$ after recovery.}
\label{x}
\end{minipage}
\end{figure}

\textit{\textbf{B.2. Derivation of Formulas}}\\
To find the probability that a variable node is resolved, we first need to characterize the set of variable nodes resolved by each algorithm. A careful inspection of the iterative algorithms under consideration and based on Theorem \ref{Model_Iter} in section \ref{analysis}, in general, the variable nodes in the set $\mathcal{R}_\mathcal{X} \triangleq \left\{\mathcal{X}_\beta\cup\mathcal{X}_{\beta+1}\cup\cdots\cup\mathcal{X}_{d_v}\right\}$ are recovered and those in the set $\mathcal{R}^{c}_\mathcal{X} \triangleq \left\{\mathcal{X}_0\cup\mathcal{X}_1\cup\cdots\cup\mathcal{X}_{\beta-1}\right\}$ are left intact, where the value $\beta$ depends on the algorithm.\\
Thus, the probability $\raisebox{2pt}{$p$}^{(\ell)}_{r}$ of a variable node in $\mathcal{K}^{(\ell)}$ being recovered is:
\begin{equation}
\raisebox{2pt}{$p$}^{(\ell)}_{r} = \sum_{\mathcal{X}_i\in\mathcal{R}_{\mathcal{X}}}{\raisebox{2pt}{$p$}^{(\ell)}_{\mathcal{X}_i}}.
	\label{eq:v}
\end{equation}
Therefore, according to the total probability theorem, the probability of a variable node $v$ remaining unresolved, i.e., $v\in\mathcal{K}^{(\ell+1)}$, is:
\[
\alpha^{(\ell+1)} = \alpha^{(\ell)}\left(1-\raisebox{2pt}{$p$}^{(\ell)}_{r}\right).
\]
When a variable node is recovered, its $d_v$ edges along with the variable node itself are removed from the subgraph induced by $\mathcal{K}^{(\ell)}$ and therefore, check nodes incident to these removed edges would face a reduction in their degree. We denote by $p^{(\ell)}_{\mathcal{N}_{ij}}$ the probability that a check node $c$ turns from degree $i$ in iteration $\ell$ (i.e. $c\in\mathcal{N}^{(\ell)}_i$) to degree $j\leq i$ in iteration $\ell+1$ (i.e. $c\in\mathcal{N}^{(\ell+1)}_j$). This happens if out of $i$ edges emanating from $c$ and incident to the set of unresolved variable nodes $\mathcal{K}^{(\ell)}$, $i-j$ of them are removed.\\
On the other side of the graph, when a variable node $v\in\mathcal{X}_i$ is recovered (i.e., $\mathcal{X}_i\subset\mathcal{R}_\mathcal{X}$), by definition, out of $d_v$ edges emanating from $v$, $i$ are connected to the set $\mathcal{N}_1$ and $d_v-i$ are connected to the set $\mathcal{R}_\mathcal{N}\triangleq\left\{\mathcal{N}_2\cup\mathcal{N}_3\cup\cdots\cup\mathcal{N}_{d_c}\right\}$.\\
In the asymptotic case, as $n$ grows large, we assume that the graph has a random structure in every iteration. Therefore, for each recovered variable node $v$, the set of $i$ and $d_v-i$ removed edges are distributed uniformly with respect to the check nodes in $\mathcal{N}_1$ and $\mathcal{R}_\mathcal{N}$, respectively. As we have two sets $\mathcal{N}_1$ and $\mathcal{R}_\mathcal{N}$ to deal with, we differentiate between $p^{(\ell)}_{\mathcal{N}_{10}}$ and $p^{(\ell)}_{\mathcal{N}_{ij}}$ ($i>1$). Once the probabilities $p^{(\ell)}_{\mathcal{N}_{10}}$ and $p^{(\ell)}_{\mathcal{N}_{ij}}$ are found, the new check node degree distribution $p_{\mathcal{N}_j}^{(\ell+1)}$ with respect to the subgraph induced by $\mathcal{K}^{(\ell+1)}$ can then be derived using the total probability law:
\[
p_{\mathcal{N}_j}^{(\ell+1)} = \sum_{i=j}^{d_c}{p_{\mathcal{N}_{i}}^{(\ell)}p_{\mathcal{N}_{ij}}^{(\ell)}},\hspace{20pt}j=0,\cdots,d_c.
\]
To find the probabilities $p^{(\ell)}_{\mathcal{N}_{10}}$ and $p^{(\ell)}_{\mathcal{N}_{ij}}$, $i>1$, we denote by $\raisebox{2pt}{$p$}_{d=1}$ and $\raisebox{2pt}{$p$}_{d>1}$ the conditional probabilities that an edge in the induced subgraph is removed given that it is incident to a check node in the set $\mathcal{N}_1$ and $\mathcal{R}_\mathcal{N}$, respectively. It then follows that:
\[
p^{(\ell)}_{\mathcal{N}_{10}} = {1\choose{1}} \left(\raisebox{2pt}{$p$}^{(\ell)}_{d=1}\right)^{1}\left(1-\raisebox{2pt}{$p$}^{(\ell)}_{d=1}\right)^0 = \raisebox{2pt}{$p$}^{(\ell)}_{d=1},\hspace{1cm}p^{(\ell)}_{\mathcal{N}_{11}} = 1-p^{(\ell)}_{\mathcal{N}_{10}}.
\]
and
\[
p^{(\ell)}_{\mathcal{N}_{ij}} = {i\choose{i-j}} \left(p^{(\ell)}_{d>1}\right)^{i-j}\left(1-p^{(\ell)}_{d>1}\right)^j,\hspace{20pt} i=2,\cdots,d_c,\hspace{20pt} j=0,\cdots,i.
\]
The probability $\raisebox{2pt}{$p$}^{(\ell)}_{d=1}$ can then be calculated as follows:
\begin{align*}
\raisebox{2pt}{$p$}^{(\ell)}_{d=1} &= \Pr[h_e\in\mathcal{R}^{(\ell)}_\mathcal{X}|t_e\in\mathcal{N}^{(\ell)}_1,h_e\in\mathcal{K}^{(\ell)}]
= \displaystyle\sum_{ \mathcal{X}_i\in\mathcal{R}_{\mathcal{X}}}{\Pr[h_e\in\mathcal{X}^{(\ell)}_i|t_e\in\mathcal{N}^{(\ell)}_1,h_e\in\mathcal{K}^{(\ell)}]},\\
&= \displaystyle\sum_{ \mathcal{X}_i\in\mathcal{R}_{\mathcal{X}}}{\frac{\Pr[t_e\in\mathcal{N}^{(\ell)}_1|h_e\in\mathcal{X}^{(\ell)}_i,h_e\in\mathcal{K}^{(\ell)}] \Pr[h_e\in\mathcal{X}^{(\ell)}_i|h_e\in\mathcal{K}^{(\ell)}]}{\Pr[t_e\in\mathcal{N}^{(\ell)}_1|h_e\in\mathcal{K}^{(\ell)}]}}
= \displaystyle\sum_{\mathcal{X}_i\in\mathcal{R}_{\mathcal{X}}}{\frac{\frac{i}{d_v}\raisebox{2pt}{$p$}^{(\ell)}_{\mathcal{X}_i}}{p^{(\ell)}}} = \displaystyle\frac{\displaystyle\sum_{\mathcal{X}_i\in\mathcal{R}_{\mathcal{X}}}{i\raisebox{2pt}{$p$}^{(\ell)}_{\mathcal{X}_i}}}{d_vp^{(\ell)}}.
\end{align*}
where, $p^{(\ell)}$ is the probability of an edge $e$ being adjacent to a check node in $\mathcal{N}^{(\ell)}_1$ conditioned on the fact that it is adjacent to a variable node in $\mathcal{K}^{(\ell)}$ (refer to Figure \ref{K1}). By using Bayes' rule, this probability is calculated as:
\begin{equation}
p^{(\ell)} = \Pr[t_e\in \mathcal{N}^{(\ell)}_1|h_e\in\mathcal{K}^{(\ell)}] = \frac{\Pr[h_e\in\mathcal{K}^{(\ell)}|t_e\in \mathcal{N}^{(\ell)}_1]\Pr[t_e\in \mathcal{N}^{(\ell)}_1]}{\Pr[h_e\in\mathcal{K}^{(\ell)}]} = \frac{\frac{1}{d_c}\times \raisebox{2pt}{$p$}^{(\ell)}_{\mathcal{N}_1}}{\alpha^{(\ell)}} = \frac{\raisebox{2pt}{$p$}^{(\ell)}_{\mathcal{N}_1}}{\alpha^{(\ell)} d_c}.
	\label{eq:p}	
\end{equation}
The probability $\raisebox{2pt}{$p$}^{(\ell)}_{d>1}$ can be computed following similar steps:
\begin{align*}
\raisebox{2pt}{$p$}^{(\ell)}_{d>1} &= \Pr[h_e\in\mathcal{R}^{(\ell)}_\mathcal{X}|t_e\in\mathcal{R}^{(\ell)}_\mathcal{N},h_e\in\mathcal{K}^{(\ell)}]
= \displaystyle\sum_{ \mathcal{X}_i\in\mathcal{R}_{\mathcal{X}}}{\Pr[h_e\in\mathcal{X}^{(\ell)}_i|t_e\in\mathcal{R}^{(\ell)}_\mathcal{N},h_e\in\mathcal{K}^{(\ell)}]}\\
&= \displaystyle\sum_{ \mathcal{X}_i\in\mathcal{R}_{\mathcal{X}}}{\frac{\Pr[t_e\in\mathcal{R}^{(\ell)}_\mathcal{N}|h_e\in\mathcal{X}^{(\ell)}_i,h_e\in\mathcal{K}^{(\ell)}] \Pr[h_e\in\mathcal{X}^{(\ell)}_i|h_e\in\mathcal{K}^{(\ell)}]}{\Pr[t_e\in\mathcal{R}^{(\ell)}_\mathcal{N}|h_e\in\mathcal{K}^{(\ell)}]}}\\
&= \displaystyle\sum_{ \mathcal{X}_i\in\mathcal{R}_{\mathcal{X}}}{\frac{\left(1-\Pr[t_e\in\mathcal{N}^{(\ell)}_1|h_e\in\mathcal{X}^{(\ell)}_i,h_e\in\mathcal{K}^{(\ell)}]\right) \Pr[h_e\in\mathcal{X}^{(\ell)}_i|h_e\in\mathcal{K}^{(\ell)}]}{\left(1-\Pr[t_e\in\mathcal{N}^{(\ell)}_1|h_e\in\mathcal{K}^{(\ell)}]\right)}}\\
&= \displaystyle\sum_{\mathcal{X}_i\in\mathcal{R}_{\mathcal{X}}}{\frac{\left(1-\frac{i}{d_v}\right)\raisebox{2pt}{$p$}^{(\ell)}_{\mathcal{X}_i}}{1-p^{(\ell)}}} = \displaystyle\frac{\displaystyle\sum_{\mathcal{X}_i\in\mathcal{R}_{\mathcal{X}}}{\raisebox{2pt}{$p$}^{(\ell)}_{\mathcal{X}_i}}-\displaystyle\sum_{ \mathcal{X}_i\in\mathcal{R}_{\mathcal{X}}}{\frac{i}{d_v}\raisebox{2pt}{$p$}^{(\ell)}_{\mathcal{X}_i}}}{1-p^{(\ell)}}\\
&= 
\frac{d_v \raisebox{2pt}{$p$}^{(\ell)}_{r}- \displaystyle\sum_{\mathcal{X}_i\in\mathcal{R}_{\mathcal{X}}}{i\raisebox{2pt}{$p$}^{(\ell)}_{\mathcal{X}_i}}}{d_v\left(1 - p^{(\ell)}\right)}.
\end{align*}
Given $p^{(\ell+1)}_{\mathcal{N}_i}$, the updated set $\mathcal{K}^{(\ell+1)}$ should be re-partitioned into the sets $\mathcal{X}^{(\ell+1)}_i$. By definition, a variable node $v$ in $\mathcal{X}_i$ has $i$ connections to $\mathcal{N}_1$ and $d_v-i$ connections to the set $\mathcal{R}_\mathcal{N}$. Therefore, if one of the adjacent check nodes of $v$ in $\mathcal{R}^{(\ell)}_\mathcal{N}$ turns to a check node in $\mathcal{N}^{(\ell+1)}_1$, $v$ will move from $\mathcal{X}^{(\ell)}_{i}$ to $\mathcal{X}^{(\ell+1)}_{i+1}$. This is shown in Figure \ref{rec}.\\
We denote by $\mathcal{N}^{(\ell)+}_1$ the set of check nodes that move from $\mathcal{R}^{(\ell)}_\mathcal{N}$ to $\mathcal{N}^{(\ell+1)}_1$. The configuration of $\mathcal{N}^{(\ell+1)}_1$, $\mathcal{N}^{(\ell)}_{11}$ and $\mathcal{N}^{(\ell)+}_1$ is depicted in Figure \ref{x}.\\
We also refer to the set of edges that have their tail in the set $\mathcal{R}^{(\ell)}_\mathcal{N}$ as \textit{free edges}. Due to the random structure of the graph assumed in the asymptotic case, edges connected to the set $\mathcal{N}^{+}_1$ are uniformly distributed with respect to free edges.\\
It can be seen that the probability $\raisebox{2pt}{$p$}^{(\ell)}_{\mathcal{X}_{ij}}$ defined as the probability of a variable node $v\in \mathcal{X}^{(\ell)}_i$ turning to $v\in\mathcal{X}^{(\ell+1)}_j$ is calculated by:
\begin{equation}
\label{pxij}
p_{\mathcal{X}_{ij}}^{(\ell)} = \Pr[v\in\mathcal{X}^{(\ell+1)}_j|v\in\mathcal{X}^{(\ell)}_i,v\in\mathcal{K}^{(\ell+1)}] = {d_v-i\choose j-i}\left(\raisebox{2pt}{$p$}^{(\ell)}_{x}\right)^{j-i}\left(1-\raisebox{2pt}{$p$}^{(\ell)}_{x}\right)^{d_v-j}, \hspace{20pt} j=i,\cdots,d_v,\hspace{10pt} i=0,\cdots,\beta-1.
\end{equation}
where $\beta$ is the algorithm dependent parameter defined in conjunction with the set $\mathcal{R}_\mathcal{X}$, and $\raisebox{2pt}{$p$}^{(\ell)}_{x}$ is defined as the probability that a free edge corresponds to the set $\mathcal{N}^{(\ell)+}_1$. Note that such an edge will have its head in the set $\mathcal{R}^c_\mathcal{X}$ because the set $\mathcal{R}_\mathcal{X}$ is completely resolved.\\
Thus, based on the total probability law we have:
\begin{equation}
\raisebox{2pt}{$p$}^{(\ell+1)}_{\mathcal{X}_j}= \frac{\displaystyle\sum_{i=0}^{\min\{j,\beta-1\}}{\raisebox{2pt}{$p$}^{(\ell)}_{\mathcal{X}_i} \raisebox{2pt}{$p$}^{(\ell)}_{\mathcal{X}_{ij}}}}{1-\raisebox{2pt}{$p$}^{(\ell)}_{r}}, \hspace{20pt} j=0,\cdots,d_v.
	\label{p_x_1}
\end{equation}
The denominator is just a normalization factor making $\raisebox{2pt}{$p$}^{(\ell+1)}_{\mathcal{X}_j}$ a valid probability measure. It is derived as:
\[
\displaystyle\sum_{j=0}^{d_v}{\raisebox{2pt}{$p$}^{(\ell+1)}_{\mathcal{X}_j}}=\sum_{ \mathcal{X}_i\in\mathcal{R}^c_{\mathcal{X}}}{\sum_{j=i}^{d_v}{\raisebox{2pt}{$p$}^{(\ell)}_{\mathcal{X}_i} \raisebox{2pt}{$p$}^{(\ell)}_{\mathcal{X}_{ij}}}} = \sum_{\mathcal{X}_i\in\mathcal{R}^c_{\mathcal{X}}}{p_{\mathcal{X}_i}^{(\ell)}} = 1-\raisebox{2pt}{$p$}^{(\ell)}_{r}.
\]
The probability $\raisebox{2pt}{$p$}^{(\ell)}_{x}$ is calculated as follows:
{\setlength\arraycolsep{1pt}
\begin{eqnarray}
\raisebox{3pt}{$p$}^{(\ell)}_x &=& \Pr[t_e\in\mathcal{N}^{(\ell)+}_1|h_e\in\mathcal{K}^{(\ell+1)},t_e\notin\mathcal{N}_{11}^{(\ell)}],\nonumber\\
&=& \frac{\raisebox{2pt}{$p$}^{(\ell)+}_{\mathcal{N}_1}}{\raisebox{2pt}{$p$}^{(\ell)+}_{\mathcal{N}_1} + \displaystyle\sum_{i=2}^{d_c}{i\raisebox{2pt}{$p$}^{(\ell+1)}_{\mathcal{N}_i}}}.
\label{p_x}
\end{eqnarray}
}
where,
\[
\raisebox{2pt}{$p$}^{(\ell)+}_{\mathcal{N}_1} = \sum_{j=2}^{d_c}{\raisebox{2pt}{$p$}^{(\ell)}_{\mathcal{N}_j} \raisebox{2pt}{$p$}^{(\ell)}_{\mathcal{N}_{j1}}}.
\]
This probability will be inserted in (\ref{pxij}) for the calculation of $\raisebox{3pt}{$p$}^{(\ell)}_{\mathcal{X}_{ij}}$.\\

\textit{\textbf{B.3. Pre-phase Iteration for Genie, XH and SBB Algorithms}}\\
The initial density factor is denoted by $\alpha^{(0)}$. In a random graph, an edge emanates from a variable node in the set $\mathcal{K}^{(0)}$ with probability $\alpha^{(0)}$. Therefore, $\raisebox{2pt}{$p$}^{(0)}_{\mathcal{N}_i}$, the probability of a check node being in the set $\mathcal{N}^{(0)}_i$, is given by the following binomial distribution.
\begin{equation}
\raisebox{2pt}{$p$}^{(0)}_{\mathcal{N}_i} = {d_c\choose i} \left(\alpha^{(0)}\right)^i \left(1-\alpha^{(0)}\right)^{d_c-i},\hspace{20pt}i=0,\cdots,d_c.
\label{p_n}
\end{equation}
To find the probability $\raisebox{2pt}{$p$}^{(0)}_{\mathcal{X}_i}$, we need the probability $p^{(0)}$ defined and calculated in equation (\ref{eq:p}). Knowing $p^{(0)}$ the probability $\raisebox{2pt}{$p$}^{(0)}_{\mathcal{X}_i}$ will follow a binomial distribution as follows:
\begin{equation}
\raisebox{2pt}{$p$}^{(0)}_{\mathcal{X}_i} = {d_v\choose i}\left(p^{{(0)}}\right)^i\left(1-p^{(0)}\right)^{d_v-i},\hspace{20pt}i=0,\cdots,d_v.
	\label{eq:x}
\end{equation}

\begin{figure}[h]
\vspace{20pt}
\centering
\includegraphics[width = 200 pt]{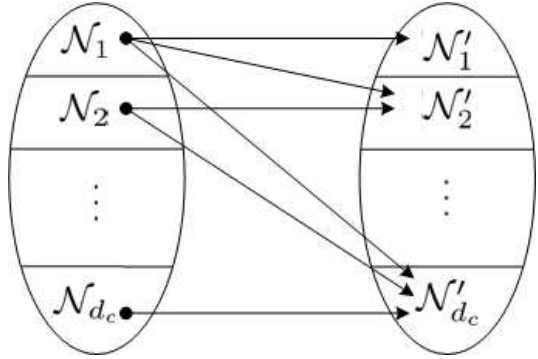}
\caption{Transition from $\mathcal{N}$ to $\mathcal{N}'$}
\label{NNpConversion}
\end{figure}

\begin{figure}[h]
\vspace{20pt}
\centering
\includegraphics[width = 250 pt]{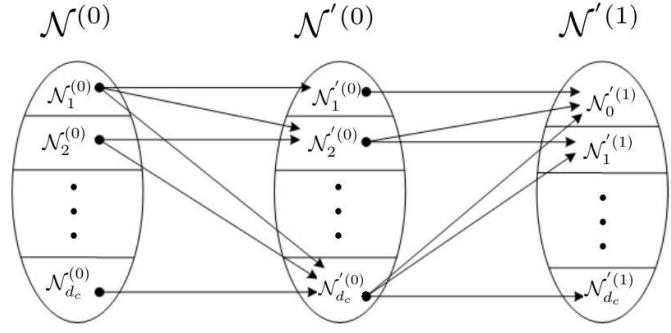}
\caption{Evolution of $\mathcal{N}$ due to $K_\Delta$ and the resolution of variable nodes in the first iteration}
\label{Dependencies}
\end{figure}

\textit{\textbf{B.4. Pre-phase Iterations for LM Algorithm}}\\
In this section we drop all the superscripts representing the iteration number for the ease of notation. It will be introduced when there is a potential ambiguity.\\
Starting from the initial density factor $\alpha^{(0)}$, probability $\raisebox{2pt}{$p$}^{(0)}_{\mathcal{N}_i}$ of check degree distribution in the induced subgraph by the set $\mathcal{K}^{(0)}$ can be calculated from (\ref{p_n}).\\
In the first iteration of the LM algorithm, the variable nodes adjacent to at least one zero-valued check node are set to zero. The set of remaining variable nodes are called \emph{potential support set} and are denoted by $\mathcal{K}'$. This set is a combination of the real support set $\mathcal{K}$ and an additional set $\mathcal{K}_\Delta$; The set of all zero-valued variable nodes that have $d_v$ connections to the nonzero check nodes $\mathcal{N}_{\neq 0}$. The probability $\raisebox{2pt}{$p$}_{\mathcal{K}_\Delta}$ that a variable node belongs to the set $\mathcal{K}_\Delta$ is calculated as:
\[
\raisebox{3pt}{$p$}_{\mathcal{K}_\Delta} = \raisebox{2pt}{$p$}^{d_v}_{\Delta}.
\]
where, $\raisebox{2pt}{$p$}_{\Delta}$ is the probability that an edge from zero-valued variable nodes terminates in $\mathcal{N}_{\neq 0}$. This probability is:
\begin{align*}
\raisebox{2pt}{$p$}_{\Delta} &= \Pr[t_e\in\mathcal{N}_{\neq 0}|h_e\notin\mathcal{K}] = 1 - \Pr[t_e\in\mathcal{N}_0|h_e\notin\mathcal{K}],\\
&= 1 - \frac{\Pr[h_e\notin\mathcal{K}|t_e\in\mathcal{N}_0]\Pr[t_e\in\mathcal{N}_0]}{\Pr[h_e\notin\mathcal{K}]} = 1 - \frac{p\raisebox{-5pt}{$\scriptstyle{\mathcal{N}_0}$}}{1-\alpha},\\
&= 1- \left(1-\alpha\right)^{d_c-1}.
\end{align*}
In this algorithm we have to group the check nodes based on the number of connections they have to the potential support set, rather than the original support set. This brings sets, denoted by $\mathcal{N}'_i$, into play that reflect the effect of $\mathcal{K}_\Delta$ on the degree distribution of check nodes. As $\mathcal{K}_\Delta$ does not change the size of $\mathcal{N}_{\neq 0}$, to calculate the probability of each subset $\mathcal{N}'_i$, we calculate the probability of each set $\mathcal{N}_i$ as before and then account for the effect of $\mathcal{K}_\Delta$ on changing the degrees. This process can be seen in Figure \ref{NNpConversion}.
With an abuse of notation, we will denote by $\raisebox{2pt}{$p$}_{\mathcal{N}_{ij}}$ the probability that a check nodes is transfered from $\mathcal{N}_i$ to $\mathcal{N}'_j$. This probability can be calculated as follows:
\[
\raisebox{3pt}{$p$}_{\mathcal{N}_{ij}} = {{d_c-i}\choose{j-i}}\left(p'\right)^{j-i}\left(1-p'\right)^{d_c - j},\hspace{20pt}i=1,\cdots,d_c,\hspace{20pt}j=i,\cdots,d_c.
\]
where, $p'$ is the probability that a free edge from $\mathcal{N}_{\neq 0}$ goes to $\mathcal{K}_\Delta$ and is calculated as:
\begin{align*}
p' &= \Pr[h_e\in\mathcal{K}_\Delta|t_e\in\mathcal{N}_{\neq 0},h_e\notin\mathcal{K}] = \frac{\Pr[t_e\in\mathcal{N}_{\neq 0}|h_e\in\mathcal{K}_\Delta]\Pr[h_e\in\mathcal{K}_\Delta|h_e\notin\mathcal{K}]}{\Pr[t_e\in\mathcal{N}_{\neq 0}|h_e\notin\mathcal{K}]} = \frac{1\times 
\raisebox{3pt}{$p$}_{\mathcal{K}_\Delta}}{\raisebox{3pt}{$p$}_{\Delta}} = \raisebox{3pt}{$p$}_{\Delta}^{d_v-1}.
\end{align*}
Thus,
\[
\raisebox{3pt}{$p$}_{\mathcal{N}'_0} = \raisebox{3pt}{$p$}_{\mathcal{N}_0}, \hspace{1cm} \raisebox{3pt}{$p$}_{\mathcal{N}'_j} = \sum_{i=1}^{j}{\raisebox{3pt}{$p$}_{\mathcal{N}_i} \raisebox{3pt}{$p$}_{\mathcal{N}_{ij}}} ,\hspace{20pt}j=1,\cdots,d_c.
\]
Variable nodes in $\mathcal{K}_\Delta$ are not connected to the set $\mathcal{N}_1'$. Thus, the support set $\mathcal{K}$ is divided into $\mathcal{X}_i$ according to equation (\ref{eq:x}). The only difference is that $\raisebox{3pt}{$p$}_{\mathcal{N}'_1}$ should be used instead of $\raisebox{2pt}{$p$}^{(0)}_{\mathcal{N}_1}$ in the calculation of $p^{(0)}$. Also, variable nodes in $\mathcal{K}_\Delta$ will contribute to $\mathcal{X}'_0$. This means that:
\[
\raisebox{2pt}{$p$}_{\mathcal{X}'_0} = \raisebox{2pt}{$p$}_{\mathcal{X}_0} + \raisebox{2pt}{$p$}_{\mathcal{K}_\Delta}.
\]
In this algorithm, like the Genie, $\beta = 1$ and therefore:
\[
\raisebox{2pt}{$p$}_{r} = \sum_{i=1}^{d_v}{\raisebox{2pt}{$p$}_{\mathcal{X}'_i}},\hspace{1cm}\alpha^{(1)}=\alpha^{(0)}\left(1-\raisebox{2pt}{$p$}_{r}\right).
\]
The pre-phase in the LM algorithm has two steps before we can use the general formulation presented in section \ref{analysis}. To find out the probabilities $\raisebox{2pt}{$p$}^{(1)}_{\mathcal{N}'}$, we use the intermediate probabilities $\raisebox{2pt}{$p$}_{\mathcal{N}'_{ij}}$ to denote the probability that a check node goes from $\mathcal{N}_i^{'(0)}$ to $\mathcal{N}_j^{'(1)}$. The complication is because in the first iteration all the released edges come from $\mathcal{K}^{(0)}$ and should not affect the connections that $\mathcal{N}_i^{'(0)}$s have with $\mathcal{K}_\Delta$. The only way to deal with this case is to go back to $\mathcal{N}_{ij}$. The general picture is depicted in Figure \ref{Dependencies}.
Note that:
\begin{enumerate}
	\item $\mathcal{N}_{ij}$ has $j-i$ connections to $\mathcal{K}_\Delta$, $i$ connections to $\mathcal{K}$ and $d_c-j$ free edges.
	\item Check nodes may change from $\mathcal{N}_j^{(0)}$ to $\mathcal{N}^{'(1)}_q$ when there are $j-q$ connections to $\mathcal{K}$; i.e., $j \geq q$.
\end{enumerate}
Thus, to go from $\mathcal{N}_i^{(0)}$ to $\mathcal{N}^{'(1)}_q$, we need to go from $\mathcal{N}_i^{(0)}$ to intermediate $\mathcal{N}_j^{'(0)}$, where $j-q< i\leq j$, and then from $\mathcal{N}_j^{'(0)}$ to $\mathcal{N}^{'(1)}_q$. Thus, the overall formula would be:
\[
\raisebox{2pt}{$p$}^{(1)}_{\mathcal{N}'_q} = \sum_{j=\max\{2,q\}}^{d_c} \sum_{i=\max\{1,j-q\}}^{j}{\raisebox{2pt}{$p$}^{(0)}_{\mathcal{N}_i} \raisebox{2pt}{$p$}^{(0)}_{\mathcal{N}_{ij}}{i\choose{j-q}}\left(\raisebox{2pt}{$p$}_{f}\right)^{j-q}\left(1-\raisebox{2pt}{$p$}_{f}\right)^{i-j+q}} ,\hspace{20pt}q=1,\cdots,d_c
\]
\[
\raisebox{2pt}{$p$}^{(1)}_{\mathcal{N}'_0} = \sum_{i=1}^{d_c}{\raisebox{2pt}{$p$}^{(0)}_{\mathcal{N}_i} \raisebox{2pt}{$p$}^{(0)}_{\mathcal{N}_{ii}}\left(\raisebox{2pt}{$p$}_{f}\right)^{i}} 
\]
where $\raisebox{2pt}{$p$}_{f}$ is as follows:
\begin{align*}
\raisebox{2pt}{$p$}_{f} &= \Pr[h_e\in\mathcal{X}_r|t_e\in\mathcal{N}_u,h_e\in\mathcal{K}]\\
	&= \frac{\Pr[t_e\in\mathcal{N}_u|h_e\in\mathcal{X}_r,h_e\in\mathcal{K}]\Pr[h_e\in\mathcal{X}_r|h_e\in\mathcal{K}]}{\Pr[t_e\in\mathcal{N}_u|h_e\in\mathcal{K}]}\\
	&= \frac{\left(1-\displaystyle\frac{\displaystyle\sum_{i=1}^{d_v}{i \raisebox{2pt}{$p$}_{\mathcal{X}'_i}}}{d_v \raisebox{2pt}{$p$}_{r}} \right)p\raisebox{-5pt}{$\scriptstyle r$}}{1-p^{(0)}}\\
	&= \frac{\raisebox{2pt}{$p$}_{r}-p^{(0)}}{1-p^{(0)}}
\end{align*}
where $\mathcal{N}_u=\left\{\{\mathcal{N}_1\cup\mathcal{N}_2\cup\cdots\cup\mathcal{N}_{d_c}\}\backslash \mathcal{N}^{'(0)}_1\right\}$.\\
From this point forth, the formulation presented in the general framework can be used with the probabilities $p_{\mathcal{X}'}$ and $p_{\mathcal{N}'}$ replacing $p_{\mathcal{X}}$ and $p_{\mathcal{N}}$.
\end{document}